# Bounded Delay Packet Scheduling in a Bounded Buffer

Stanley P. Y. Fung [*]

October 31, 2018

## Abstract

We study the problem of buffer management in QoS-enabled network switches in the bounded delay model where each packet is associated with a weight and a deadline. We consider the more realistic situation where the network switch has a finite buffer size. A 9.82-competitive algorithm is known for the case of multiple buffers (Azar and Levy, SWAT'06). Recently, for the case of a single buffer, a 3-competitive deterministic algorithm and a 2.618-competitive randomized algorithm was known (Li, INFOCOM'09). In this paper we give a simple deterministic 2-competitive algorithm for the case of a single buffer.

## 1 Introduction

In the *bounded delay buffer management problem*, introduced in [9], we have a set of packets arriving at at network switch, each having a release time, a deadline, and a weight. The weight represents a Quality of Service (QoS) parameter, i.e. how important it is to send the packet. A packet beyond its deadline will lose its value. The objective is to send a packet at each time step so that the total weight of sent packets is maximized. This can be modelled as a *unit job scheduling* problem, and an extensive line of work [15, 4, 6, 13] has been done in the past few years on the problem. A simple greedy algorithm which always sends the heaviest packet is 2-competitive [8, 9]. The current best deterministic results are a 1.828 upper bound [6] and a 1.618 lower bound [5, 8, 15].

In all the above work, the buffer is assumed to have an infinite capacity. This is of course not true in practice. The addition of finite buffer size makes the problem no longer directly equivalent to the unit job scheduling problem. It would be useful to investigate the design and analysis of algorithms for the bounded buffer case.

There is another line of work on the *FIFO buffer management problem*, where the buffer is of finite size, but instead of a deadline, packets must be sent in a FIFO manner (i.e. leave the switch in the same order as they arrive, although packets can be dropped). Both the case of a single buffer [9, 10, 14, 7] and multiple buffers [3, 2] have been studied.

Returning to the model of bounded delay model with bounded buffers, a multiple-buffer version of the problem was considered in [1]. In this setting there are $m \geq 1$ buffers, each with a finite size. Each packet arrives to a specific buffer, and at each time step one packet can be sent from one of the buffers. The paper gave a 9.82-competitive algorithm for the problem. Recently, Li [11, 12] gave several results, including a 3-competitive deterministic algorithm and a 2.618-competitive randomized algorithm for the single-buffer case, and a 4.723-competitive algorithm for the multiple-buffer case.

---

[*]Department of Computer Science, University of Leicester, Leicester, United Kingdom. pyfung@mcs.le.ac.uk

In this paper we give a simple algorithm GREEDYQUEUE (GRQ), for the single buffer case, and show that it is 2-competitive. It is therefore even better than the randomized algorithm in [11].

## 2 The GRQ Algorithm

The simple greedy algorithm (that always keeps the heaviest packets in the buffer, and sends the heaviest one at each time step) is not competitive in the bounded buffer case, even for a single buffer, for the simple reason that it holds up precious buffer space for packets that cannot be sent anyway. For example if there are a lot of packets with deadline 1 and weight 1, and other packets with long deadlines and weight $1 - \epsilon$, then the greedy algorithm will keep all weight-1 packets, all except one will expire after one time step.

However, we show that a simple modification of the greedy idea gives a 2-competitive algorithm. The idea is to keep track of a 'provisional schedule' and prevent packets that cannot be sent anyway from occupying the buffer. Our provisional schedule is not the same as the 'optimal provisional schedule' in [12] which maximizes the total value of packets assuming no future packet arrivals. In fact they showed that, using such optimal provisional schedules for packet admission together with greedy algorithm for packet transmission does not give better than 4-competitiveness. More generally, any algorithm using such optimal provisional schedules for packet admission decisions cannot be better than 2-competitive. Our provisional schedule is designed to work closely with the greedy algorithm, in that it will not store packets that the optimal schedule keeps but the greedy algorithm does not.

**Notation.** Time is divided into discrete time steps. Each time step is divided into the *arrival stage* and the *transmission stage*. In the arrival stage, a set of packets arrive to the buffer. Let $B$ denotes the size of the buffer, i.e. it can hold at most $B$ packets. If the total number of current and new packets exceeds $B$, some packets must be discarded so that the total number remains at most $B$. Then at the transmission stage, at most one packet will be sent and removed from the buffer. Each packet $p$ has release time $r(p)$, deadline $d(p)$ and weight $w(p)$. A packet with deadline $d(p)$ must be sent on or before time step $d(p)$ or else its value will be lost and the packet can be discarded.

We will use the following *slot naming convention*. By a 'slot' we mean a position in the buffer, or sometimes a time step (which can be seen to be equivalent below). We assign a label to each buffer position as follows. At time 1 the slots are labelled $Q[1..B]$. At time 2, the position that becomes empty after a packet is sent at time 1 is labelled $B + 1$, so now the slots are labelled $Q[2..B+1]$. In general, at any time $t$ the slots are labelled $Q[t..t + B - 1]$.

Let OPT denote the optimal offline algorithm.

**Description of the algorithm.** The algorithm arranges packets in its buffer in a queue $Q$ with $B$ slots. At time $t$, the slots are labelled $Q[t..t + B - 1]$ as described above. Slot $i$ will only hold packets will deadlines on or after $i$. The newly arrived packets together with the current packets in the buffer are considered one by one, in descending order of weights (ties broken arbitrarily), to be put into the smallest-indexed available slot in $Q$. If no such slot exists because of the packet's deadline (recall that a slot only holds packets will deadline larger than or equal to its label), the packet is discarded. After the buffer is full, any remaining packets are also discarded. When a packet is to be transmitted, the packet in the smallest-indexed slot in the queue (which is always the heaviest packet in the buffer) is sent.

**Simple properties.** We first prove the following properties.

**Lemma 1** *The weight of the packet in a slot in $Q$ never decreases, i.e., if $Q[t]$ contains a packet $p$ at some time $t_0 \leq t$, then from time $t_0$ onwards the same slot is always occupied by packets of weight at least $w(p)$, until after time $t$ when the slot no longer exists under our labelling convention.*

*Proof.* In the transmission stage no slots have their packets changed except the one with the packet sent. In the arrival stage we can imagine each newly arriving packet being inserted to a particular position in $Q$, according to its weight; this shifts other packets backwards (i.e. towards larger-indexed slots) in $Q$. This can only increase the value of packets in those slots because the packets are arranged in decreasing order of weight. If a packet is dropped in this process because of its deadline, all other packets after it in $Q$ remains unchanged. □

**Lemma 2** *If at time $t$, OPT sends a packet $x$ and GRQ sends a packet $y$ where $w(x) > w(y)$, and $x$ is not sent by GRQ before time $t$, then $x$ must be rejected by GRQ at a time $t_0$ where $t_0 + B \leq t$. Furthermore, on or after time $t_0$, all packets in $Q[t_0..t_0 + B - 1]$ have weight at least $w(x)$.*

*Proof.* If $x$ is still in GRQ's buffer at time $t$, then GRQ should schedule $x$ (or some other heavier packets) instead of $y$ at time $t$. Hence, $x$ is either already sent, or it is rejected at some time $t_0 < t$. Suppose $x$ is rejected at $t_0$ and further suppose $t_0 + B > t$. Then $Q[t]$ exists at time $t_0$ according to our labelling convention. Let $y'$ be the packet scheduled in $Q[t]$ at time $t_0$. Then since the weight of the packet at this slot does not decrease over time (Lemma 1), $w(y') \leq w(y)$. So $w(x) > w(y')$. Since $d(x) \geq t$, GRQ would place $x$ (or some other packets of weight at least $w(x)$) instead of $y'$ in slot $Q[t]$ at time $t_0$, rather than rejecting it. This is a contradiction. Hence $t_0 + B \leq t$.

At time $t_0$ all packets in the buffer must have weight at least $w(x)$ since otherwise $x$ would not be rejected (with the fact that the deadline of $x$ is at least $t$ and $t_0 + B \leq t$). This holds for all later times by Lemma 1. □

**Charging Scheme.** Next, we describe a charging scheme for mapping the packets in OPT to that in GRQ. For each time $t$, let $x$ and $y$ be the packet sent by OPT and GRQ respectively. If $x$ is sent by GRQ before time $t$, map $x$ in OPT to $x$ in GRQ in that earlier time step. These are called *S-charges* (self). Otherwise, if $w(x) \leq w(y)$, map $x$ in OPT to $y$ in GRQ. These are called *D-charges* (downward). Otherwise, we know that $x$ must be rejected by GRQ at some earlier time. Map $x$ in OPT to some packet in GRQ before time $t$ as described below. They are called *F-charges* (forward).

We construct these F-charges incrementally by imagining we are traversing in time. As each time step $t$ is traversed, new F-charges may be generated by those packets appearing in OPT at some later time steps (i.e., $> t$) but rejected by GRQ at this time step. If no such F-charges are generated we simply go to the next time step. Otherwise they are processed as follows. In order of appearance in OPT's schedule, each such F-charge is mapped to the earliest slot on or after $t$ in GRQ that has not been assigned two charges. Hence by construction, each slot receives at most two charges (it can be two F-charges, one F-charge and one D-charge, one F-charge and one S-charge, or one D-charge and one S-charge). Mappings once made are not changed.

We show by induction on time $t$ that each F-charge created at time $t$ is mapped to a slot on or before $t + B - 1$ (and since they are rejected by OPT on or after time $t + B$ by Lemma 2, they are indeed 'forward' charges). The claim is trivially true when $t = 0$.

Let $F_t$ be the set of F-charges created due to packets rejected by GRQ at time $t$. We consider the packets scheduled by GRQ in the time interval $[t..t + B - 1]$ and the charges they are going to receive. let $D_t$ be the set of packets in $[t..t + B - 1]$ in GRQ that receive a D-charge (which are made by packets in the same slots in OPT). Let $S_t$ be the set of packets in $[t..t + B - 1]$ in GRQ that receive an S-charge. We split $S_t$ into two parts $S_t^1$ and $S_t^2$, where packets in $S_t^1$ are scheduled in OPT during $[t..t + B - 1]$, and packets in $S_t^2$ are scheduled in OPT after time $t + B - 1$. Let $G_t$ be the subset of F-charges in $\cup_{i=1}^{t-1} F_i$ that were mapped to slots on or after $t$ (when we processed previous time steps). By induction hypothesis, they must originate from time steps after $t$ in OPT. We similarly split $G_t$ into $G_t^1$ and $G_t^2$. We have $|G_t^1| + |D_t| + |S_t^1| \leq B$ since these packets are scheduled in OPT during $[t..t + B - 1]$. We also have $|G_t^2| + |F_t| + |S_t^2| \leq B$ since all these packets have arrived on or before time $t + B - 1$ and still remain in OPT's buffer after time $t + B - 1$. Adding the two inequalities, we have $(|G_t| + |F_t| + |D_t| + |S_t|)/2 \leq B$. The last (i.e. $|F_t|$-th) packet in $F_t$ is mapped to slot $t - 1 + \lceil (|G_t| + |F_t| + |D_t| + |S_t|)/2 \rceil$ at the latest, because each slot can receive up to two charges and F-charges are assigned to the earliest available slot. Hence, the last (and hence all) F-charge generated at time $t$ is mapped to slot $t + B - 1$ at the latest.

The slots that receive an F-charge from $p$ must contain a packet of weight at least $w(p)$ at the time it is rejected and will remain so later (Lemma 2), and hence the F-charge on any packet in GRQ is at most its own weight.

The competitiveness now follows easily: each packet in GRQ receives at most two charges, each of weight at most that of the packet itself, so we conclude that GRQ is 2-competitive.

**Theorem 1** *GRQ is 2-competitive.*

## 3 Conclusion

Unfortunately the idea of this algorithm does not extend to the case of multiple buffers straightforwardly. The following might be a reasonable adaptation: arrange packets in descending order of weights into a 'super-queue' comprising packets from all buffers, with packets rejected if no slots are available before its deadline or if its own buffer is full. However, we can show that this algorithm has an unbounded competitive ratio. One problem is that a new packet arriving in a buffer can cause two packets to be preempted, one in the same buffer as the arriving packet and one in another buffer.

Note that our algorithm is not based on optimal provisional schedules, and therefore the lower bound of 2 in [12] does not apply here. Closing the bound [1.618, 2] is an interesting problem.